\documentclass{article}

\usepackage{arxiv}

\usepackage[utf8]{inputenc} % allow utf-8 input
\usepackage[T1]{fontenc}    % use 8-bit T1 fonts
\usepackage{url}            % simple URL typesetting
\usepackage{booktabs}       % professional-quality tables
\usepackage{amsfonts}       % blackboard math symbols
\usepackage{nicefrac}       % compact symbols for 1/2, etc.
\usepackage{microtype}      % microtypography

\usepackage{calc}
\usepackage{fancyhdr}
\usepackage{graphicx,epstopdf}
\usepackage{lastpage}
\usepackage{ifthen}
\usepackage{lineno}
\usepackage{float}
\usepackage{amsmath}
\usepackage{setspace}
\usepackage{enumitem}
\usepackage{mathpazo}
\usepackage{booktabs} % For \toprule etc. in tables
\usepackage[largestsep]{titlesec}
\usepackage{etoolbox} % For \AtBeginDocument etc.
\usepackage{tabto} % To use tab for alignment on first page
\usepackage{multirow}

\usepackage{natbib}

\usepackage{hyperref}       % hyperlinks

\begin{document}
\title{Habitable Zones in Binary Star Systems: A Zoology}

% Author Orchid ID: enter ID or remove command
%\newcommand{\orcidauthorA}{0000-0002-1398-6302} % Add \orcidA{} behind the author's name
%\newcommand{\orcidauthorB}{0000-0000-000-000X} % Add \orcidB{} behind the author's name

% Authors, for the paper (add full first names)
\author{Siegfried Eggl,\\
Rubin Observatory / \\ Department of Astronomy, \\
University of Washington, \\
 Seattle, 98015 WA, USA \\ \texttt{eggl@uw.edu} \And
Nikolaos Georgakarakos \\
Division of Science,\\ New York University Abu Dhabi,\\
Abu Dhabi, P.O. BOX 129188, UAE \\ \texttt{ng53@nyu.edu}\And
Elke Pilat-Lohinger\\
Institute for Astrophysics, \\University of Vienna, \\ 1180 Vienna, Austria\\
 \texttt{elke.pilat-lohinger@univie.ac.at}}
% Authors, for metadata in PDF
%\authorNames{Siegfried Eggl, Nikolaos Georgakarakos and Elke Pilat-Lohinger}

% Affiliations / Addresses (Add [1] after \address if there is only one affiliation.)
% \address{%
% $^{1}$ \quad LSST / DiRAC Institute, Department of Astronomy, University of Washington, Seattle, 98015 WA, USA ; eggl@uw.edu\\
% $^{2}$ \quad Division of Science, New York University Abu Dhabi, Abu Dhabi, P.O. BOX 129188, UAE; ng53@nyu.edu\\
% $^{3}$ \quad Insitute for Astrophysics, University of Vienna, 1180 Vienna, Austria; elke.pilat-lohinger@univie.ac.at}

% Contact information of the corresponding author
% \corres{Correspondence: eggl@uw.edu}
%Tel.: (optional; include country code; if there are multiple corresponding authors, add author initials) +xx-xxxx-xxx-xxxx (F.L.)}

% Current address and/or shared authorship
%\firstnote{Current address: Affiliation 3}
%\secondnote{These authors contributed equally to this work.}
% The commands \thirdnote{} till \eighthnote{} are available for further notes

%\simplesumm{} % Simple summary

%\conference{} % An extended version of a conference paper

% Abstract (Do not insert blank lines, i.e. \\)

% Keywords
\keywords{Binary Stars; Exoplanets; Habitable Zone; Astrobiology }

% The fields PACS, MSC, and JEL may be left empty or commented out if not applicable
%\PACS{J0101}
%\MSC{}
%\JEL{}

\maketitle

%%%%%%%%%%%%%%%%%%%%%%%%%%%%%%%%%%%%%%%%%%
%\setcounter{section}{-1} %% Remove this when starting to work on the template.

\begin{abstract}
Several concepts have been brought forward to determine where terrestrial planets are likely
to remain habitable in multi-stellar environments. Isophote-based habitable zones, for instance, rely on insolation geometry to predict habitability, whereas Radiative Habitable Zones take the orbital motion of a potentially habitable planet into account. Dynamically Informed Habitable Zones include gravitational perturbations on planetary orbits, and full scale, self consistent simulations promise detailed insights into the evolution of select terrestrial worlds. All of the above approaches agree that stellar multiplicity does not preclude habitability. Predictions on where to look for habitable worlds in such
environments can differ between concepts.
The aim of this article is to provide an overview of current approaches and present simple analytic estimates
for the various types of habitable zones in binary star systems.
\end{abstract}

\section{Introduction}
Defining habitable zones in binary star systems has its challenges.
Sharing a system with two host stars means that the amount and spectral composition of the light arriving at
a potentially habitable planet can vary on relatively short timescales
\citep[][]{eggl-et-al-2012, kane-hinkel-2013,  haghighipour-2013, kaltenegger-2013, cuntz-2014, forgan-2016}. The multitude of interactions between binary stars and their potentially
habitable worlds has led to the development of a variety of approaches that tackle the problem where one may find habitable worlds in such environments.
Here, we discuss a subset of the various concepts used to determine habitable zones in binary star systems found in current literature.
%Over the years, several concepts have evolved to determine where terrestrial planets are likely to retain liquid water in multi-stellar environments.
For the sake of transparency, we shall group the various approaches into the following~categories:
\begin{itemize}[leftmargin=*,labelsep=5.8mm]
\item single star habitable zones
\item isophote-based habitable zones
\item radiative habitable zones
\item dynamically informed habitable zones
\item self consistent habitable zones
\end{itemize}

The remainder of this article is dedicated to discussing the various approaches and provide analytic approximations to
calculate them where possible.

\section{Single Star Habitable Zones}
One of the first to introduce the concept of the Habitable Zone (HZ) was \citet{huang-1959}, who,
under~the assumption that the amount of radiative flux (insolation) is the main driver for climate evolution, proposed that a planet should be habitable if it orbits
its host star at a distance
\begin{equation}
r= \sqrt{\frac{L}{S}}
\label{eggl:eq:huang3}
\end{equation}
where $L$ is the luminosity of the star cast into a steradian ($L_{\odot}\;$sr), $S$ is the insolation in solar constants ($L_{\odot}\;\text{au}^{-2}\;\text{sr} = s_\odot \approx$ 1361 W m$^{-2} $)
and $r$ the distance between the planet and its host star measured in astronomical units ($\text{au}$).
Equation (\ref{eggl:eq:huang3}) states that for a sun-like star ($L=1\;L_{\odot}$) a planet receives one solar constant ($S=1\;L_{\odot}\;\text{au}^{-2}\;\text{sr}$) if it is on a circular orbit with a
semi-major axis of one astronomical unit ($r=1\;\text{au}$).
The Earth itself, however, is not on a perfectly circular orbit. Together with slight changes in the sun's luminosity this fact causes variations in the
amount of light our planet receives. Since~the Earth is still habitable, it is not unreasonable to think that the Earth's climate remains robust for a range of higher ($S_I$) and lower ($S_O$) insolation values.
Hence, the single star habitable zone (SSHZ) can be defined as a circumstellar shell between
\begin{equation}
r_I=\sqrt{\frac{L}{S_I}}\qquad \text{and} \qquad r_O=\sqrt{\frac{L}{S_O}}.
\label{eggl:eq:SSHZ}
\end{equation}
The subscripts $I$ and $O$ stand for the inner and outer
limit of the habitable zone. \citet{huang-1959, huang-1960} proposed that a terrestrial planet could remain habitable for insolation values between $S_I=5\;s_\odot$ and $S_O=0.1\;s_\odot$,
from~well inside Venus' orbit all the way to the asteroid belt.
Instead of using best guesses for habitable zone insolation limits later works  used the principle that the presence of liquid water on the surface of a planet can regulate its climate \citep{rasool-de-bergh-1970, hart-1978,hart-1979, kasting-1988, kasting-et-al-1993}. The borders of the habitable zone could then be defined as follows: Too much insolation and oceans evaporate leading to a steep increase in surface temperatures. Too little insolation and the planet falls into a cold-trap.
To calculate habitable zones around main sequence stars \citet{kasting-et-al-1993} determined effective insolation thresholds ($S_{I,O}$) that would lead to climatic runaway states.
Once found, those insolation limits could be translated to habitable zone borders using Equation (\ref{eggl:eq:SSHZ}).
\citet{kasting-et-al-1993} also discovered that not only the magnitude of radiative flux but also its spectral distribution is relevant to planetary climates.
\citet{kasting-et-al-1993} found that given the same amount of insolation energy, light originating from an M-class star is more potent
in heating an Earth-like world than that of an F-class star.
Given the effective temperature of a star ($T_{eff}$), its spectral energy distribution can often be approximated with an energy equivalent black body emission profile.
The insolation thresholds $S_{I,O}$ have, therefore, been parametrized as a function of $T_{eff}$ of the star
\citep{kopparapu-2014}
% \begin{eqnarray}
% S_I&=&1.107+1.332 \times 10^{-4}T+1.58 \times10^{-8}T^2 \label{eggl:eq:kop1}\\
%      && -8.308 \times10^{-12}T^3-1.931 \times10^{-15}T^4\nonumber \\
% S_O&=&0.356+6.171\times 10^{-5}T+1.698 \times10^{-9}T^2 \nonumber\\
%   &&-3.198 \times10^{-12}T^3-5.575 \times10^{-16}T^4, \nonumber
% \end{eqnarray}
%
\begin{eqnarray}
S_{I}&=&\text{a}_{I} + \text{b}_{I}\;T+\text{c}_{I}\ T^2+\text{d}_{I}\;T^3+\text{e}_{I}\;T^4, \label{eggl:eq:kop1a} \\
S_{O}&=&\text{a}_{O} + \text{b}_{O}\;T+\text{c}_{O}\ T^2+\text{d}_{O}\;T^3+\text{e}_{O}\;T^4. \label{eggl:eq:kop1b}
\end{eqnarray}
 $S_{I,O}$ represent insolation values corresponding to the runaway greenhouse ($S_I$) and the maximum greenhouse $(S_O)$ atmospheric collapse limits. $T=T_{eff}-5780\;\text{K}$. The coefficients $a_{I,O}$ - $e_{I,O}$ are given in Table \ref{tab:kopparapu_coeff}.
Inserting $S_{I,O}$ into Equation (\ref{eggl:eq:SSHZ}) yields the single star habitable zone borders.
The above insolation thresholds, or habitable flux limits $S_{I,O}$ contain information on the impact of the spectral distribution of the incident light as well as
the amount of light necessary to trigger a runaway state.
Single star habitable zones have been used to approximate circumstellar habitable zones in binary systems, where the planet orbits one of the stars (S-type configurations).
For binaries with orbital pericenter distances beyond 20 au this approach can yield reasonable results as we shall see in the next sections.

\begin{table}[h!]
    \centering
    \begin{tabular}{llll}
         &Inner HZ limit (I) & Outer HZ limit (O)  & units\\
        \hline
        a & 1.107                   &  0.356             & $s_\odot$\\
        b & 1.332 $\times 10^{-4}$    &6.171$\times 10^{-5}$ &$s_\odot\; \text{K}^{-1}$\\
        c & 1.580 $\times 10^{-8}$     &1.698 $\times10^{-9}$ &$s_\odot\; \text{K}^{-2}$\\
        d & -8.308 $\times10^{-12}$  &-3.198 $\times10^{-12}$&$s_\odot\; \text{K}^{-3}$\\
        e & 1.931 $\times10^{-15}$     &-5.575 $\times10^{-16}$ &$s_\odot\; \text{K}^{-4}$\\
    \end{tabular}
    \caption{Coefficients for the polynomial fit in equations (\ref{eggl:eq:kop1a}-\ref{eggl:eq:kop1b}), from \citet{kopparapu-2014}.  }
    \label{tab:kopparapu_coeff}
\end{table}

% ##################################
\section{The Trouble with Two Stars}
\label{eggl:sec:doublestars}
One implicit assumption that goes into calculations of classical habitable zone limits for single star systems as given by \citet{kasting-et-al-1993} and \citet{kopparapu-2014} is that potentially habitable worlds move about their host stars on circular orbits.
%\footnote{That is not to say that eccentric planetary motion has not been studied with respect to habitability, see e.g. \cite[][]{dressing-2010,williams-pollard-2012,kane-gelino-2012,bolmont-2016,kane-2017,way-2017,2017ApJ...837L...1M,2018AJ....155..266D,2019AJ....157..189A,2020arXiv200414673G}}.
In single star systems circular orbital motion traces stellar isophotes, i.e., regions of equal insolation around the host star. This is convenient, because it means that insolation can be considered constant over time. Near constant insolation is a good approximation in the case of the Earth, as the value of our 'solar constant' changes very little at present \citep{laskar-2004}.  Exoplanetary system architectures are diverse and can differ from the one of our Solar System, however. Many authors have investigated the effect of planetary eccentricity, and hence variable insolation, on potentially habitable worlds \citep[e.g., ][]{dressing-2010,williams-pollard-2002,kane-gelino-2012,bolmont-2016,kane-2017,way-2017,2017ApJ...837L...1M,2018AJ....155..266D,2019AJ....157..189A,2020arXiv200414673G}.
In such cases the assumption of constant insolation can break down entirely.

Similar situations arise when dealing with binary star systems where the~distances between the planet and the stars change continuously.
This means that the amount of starlight received by the planet can vary substantially.
The situation is often more complex than having a single planet on an eccentric orbit around a single star because the variations of the planetary orbital elements can occur on different timescales.
\citet{huang-1960} realized this in his study on binary star systems published not long after his work on single star habitable zones. Huang also understood that, on top of a second source of radiation, the system has to allow for stable orbital motion\footnote{We define ``stable'' orbital motion in the sense of \citet{holman-wiegert-1999}, i.e., a planet remains bound to the binary star system for a given time.} of the planet to be considered habitable.
Being expelled from the binary star system is likely not conducive to planetary habitability.
Since then, a substantial body of work has been dedicated to investigating the stability of planets in binary star systems
\citep[e.g.,][]{dvorak-1986,rabl-dvorak-1988, whitmire-et-al-1998, mardling-aarseth-2001,
holman-wiegert-1999, pilat-lohinger-dvorak-2002, pilat-lohinger-et-al-2003, pichardo-et-al-2005, doolin-blundell-2011, jaime-et-al-2012, georgakarakos-2013, quarles-2016, quarles-2018a, quarles-2018b, quarles-2018c,quarles-2020}).
Among many other important results, it has been found that stable orbits are possible in the vicinity of circumstellar and circumbinary habitable zones, but that much depends on the exact setup of the systems involved.
To have an approximate idea on where one expects stable and unstable systems one can resort to numerically generated fit functions \citep{dvorak-1986,rabl-dvorak-1988,holman-wiegert-1999,
mardling-aarseth-2001, quarles-2016}.
One~example of such an orbital stability fit function is reproduced here \citep{holman-wiegert-1999}.
For planets orbiting a star in an S-type configuration, the stability limit in terms of the semi-major axis of the planet
is given by
\begin{equation}
a_{p} < a_b \;(\text{j}+\text{k}\;\mu+\text{l}\; e_b+\text{m}\; \mu e_b+ \text{n}\; e_b^2 +\text{o}\; \mu e_b^2),
\label{eggl:eq:stabs}
\end{equation}
where $a_{p}$ is the critical initial semi-major axis of the planetary orbit, $a_b$ and $e_b$ the semi-major axis and eccentricity of the binary star orbit and $\mu=m_B/(m_A+m_B)$ the stellar mass ratio. Here, A is the star that hosts the planet and B is the perturbing star.
For a circumbinary planet we find
\begin{equation}
a_{p} > a_b\;(\text{p}+\text{q}\; e_b+\text{r}\; e_b^2+\text{s}\; \mu+\text{t}\; e_b\; \mu+\text{u}\;\mu^2+\text{v}\;e_b^2\;\mu^2).
\label{eggl:eq:stabp}
\end{equation}
The coefficients j-v and their respective uncertainties for circumstellar and circumbinary orbits are given in Table
 \ref{tab:hw99}.
Due to the vast parameter space of multi-body systems generic stability studies often use mass-less test-particles as proxies for planets. Given the intricacies of resonant interactions in N-body systems detailed numerical simulations are always recommended in order to test where a particular system permits stable orbits \citep{wittenmyer-2012, makarov-2013}.
Later studies have shown that even planets on stable, initially circular orbits may become uninhabitable due to the gravitational perturbations the two stars cause in the orbit of the planet \citep[e.g., ][]{eggl-et-al-2012, andrade-2017, quarles-2018a, quarles-2018b}.
The complex interaction of gravity and stellar radiation in binary star systems is what makes the definition of habitable zones challenging.

\begin{table}[]
    \centering
    \begin{tabular}{ccc|ccc}
        & \multicolumn{2}{c}{circumstellar} & & \multicolumn{2}{c}{circumbinary} \\
        & mean & $\pm$ & & mean & $\pm$ \\
        \hline
        j & 0.464 &0.006    &p    & 1.60 & 0.04 \\
        k & -0.380 & 0.010  &q    & 5.10 & 0.05 \\
        l & -0.631 & 0.034  &r    & -2.22 & 0.11 \\
        m & 0.586 & 0.061   &s    &  4.12 & 0.09 \\
        n & 0.150 & 0.041    &t    &  -4.27 & 0.17 \\
        o & -0.198 & 0.074  &u    & -5.09 & 0.11 \\
        -  &    -     &  -      &v    & 4.61 & 0.36 \\
        \hline
    \end{tabular}
    \caption{Orbital stability fit parameters and their respective uncertainties as given in \citet{holman-wiegert-1999} for prograde circumstellar and circumbinary motion of a massless test-planet.}
    \label{tab:hw99}
\end{table}

% ##################################
\subsection*{Isophote Based Habitable Zones}
\label{eggl:sec:ihz}
\begin{figure}[H]
\centering \includegraphics[scale=0.9]{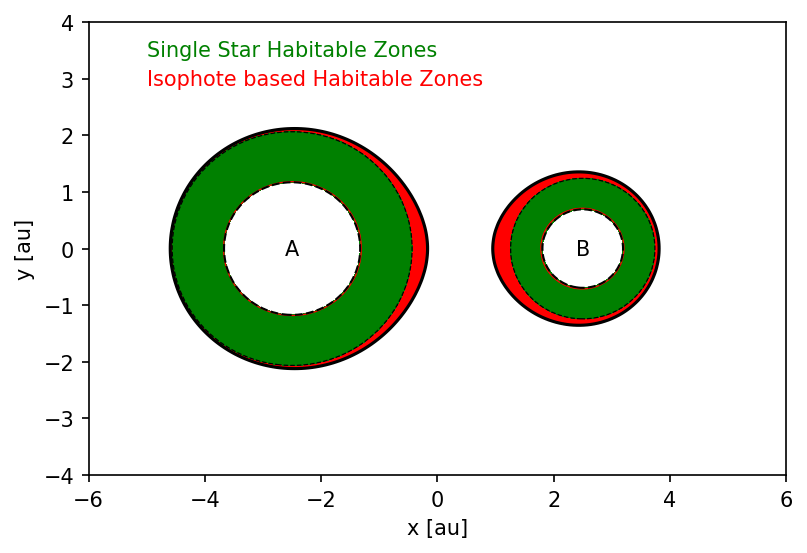}
\caption{A binary star system with circumstellar habitable zones similar to $\alpha$ Centauri on a compact orbit. The plot shows the system at a distance of $a_b=5$ au.
%  The amount of radiation a planet receives at any point in the system is
% color-coded.
Single star habitable zones (green) are shown on top of the larger isophote-based habitable zones (red).
%The white dashed lines are single star HZs.
%Planetary orbits inside of the purple dashed-dotted circles are dynamically stable.
\label{eggl:fig:stype-iso}}
\end{figure}

\begin{figure}[H]
\centering
%\begin{tabular}{ll}
\includegraphics[width=0.45\textwidth]{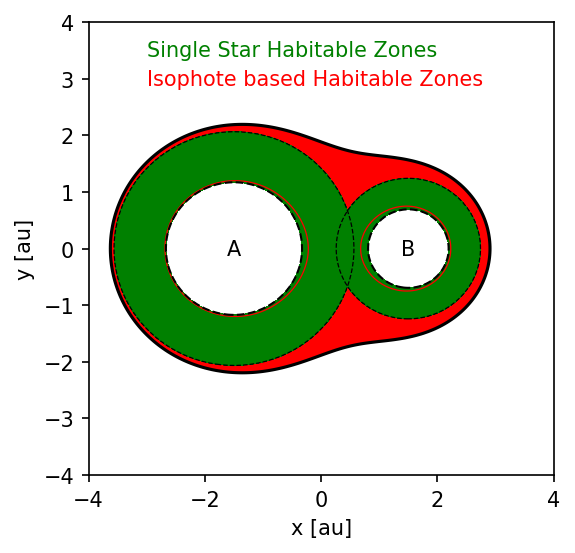}
\includegraphics[width=0.45\textwidth]{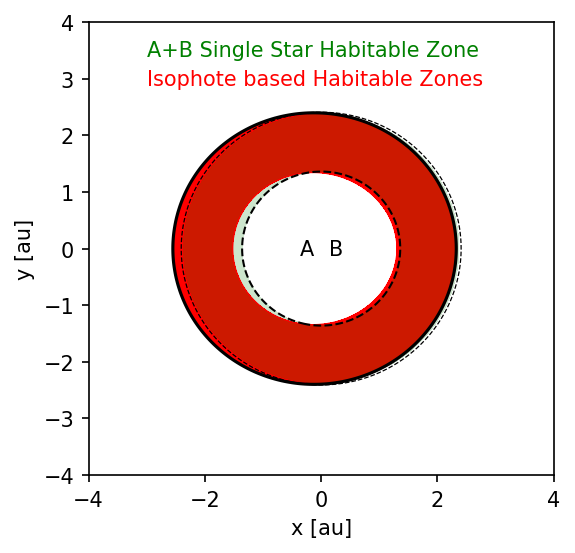}
%\end{tabular}
\caption{
\textls[-5]{Same as Figure \ref{eggl:fig:stype-iso} only for systems with semi-major axes of  $a_b=3$ au  and  $a_b=0.5$ au, respectively.}}
\label{eggl:fig:ptype-iso}       % Give a unique label
\end{figure}

Orbital dynamics aside we can define regions in binary star systems that would yield the correct amount of insolation to make a planet habitable. As we have two stars in the system, contour curves of insolation are not necessarily circular with one star at the center. Instead, the insolation geometry in the system has to be investigated in detail in order to derive isophote-based habitable zones \citep{kane-hinkel-2013, haghighipour-2013, kaltenegger-2013, mueller-haghighipour-2014}.
Figures \ref{eggl:fig:stype-iso} and \ref{eggl:fig:ptype-iso} show the instantaneous insolation values in S-type binary star systems akin to $\alpha$~Centauri ($\alpha$ Cen A, B) on close, circular orbits. Stellar parameters for $\alpha$ Cen A and B are given in Table \ref{eggl:tab:stars}.
% \ref{eggl:tab:stars} for details on the stellar parameters.
The continuous black lines in Figure \ref{eggl:fig:stype-iso} trace lines of constant insolation (isophotes).
The four curves are drawn using spectrally weighted insolation values that correspond to the inner and outer habitable zone limits for each star.
We shall refer to the area around each star enclosed by those isophotes as ``isophote-based habitable zone'' (IHZ).

\unskip

If the stars have different spectral types we also have to account for the spectral energy distribution of the combined insolation.
\citet{kane-hinkel-2013} proposed to solve this issue by superimposing the spectra of both stars weighted with the
respective star-to-planet distances. Several other studies such as
\citet{eggl-et-al-2012, haghighipour-2013, kaltenegger-2013, cuntz-2014, cuntz-2015} have adopted a slightly different approach. Instead of combining the spectra and investigating the impact
on the planetary atmosphere each star is assumed to heat the potentially habitable world independently.
If we assume that star A and B have the same spectral energy distribution each star contributes a flux inversely proportional to the square of its distance to the planet. Let $a$ and $b$ be the distances between the planet and star A and the planet and star B, respectively.
We can sum up both stellar flux contributions on the planet and require that
\begin{equation}
\frac{L_{A}}{a_I^2}+\frac{L_B}{b_I^2} \le S_I \qquad \text{and} \qquad \frac{L_A}{a_O^2}+\frac{L_B}{b^2_{O}} \ge S_O
\label{eggl:eq:shz}
\end{equation}
to guarantee a planet is habitable.
Luminosities per unit area $L_{A,B}$ are approximately constant for stars as long as they are on the main sequence.
The~distances $a$ and $b$, however, are not. Similar to the implicit equation for an ellipse $(x/u)^2+(y/w)^2=1$, Equation (\ref{eggl:eq:shz}) trace the inner (I) and outer (O) habitable zone borders when left and right hand sides are equal.
The underlying assumption that both stars have to have the same spectral energy distribution is a major shortcoming of Equation (\ref{eggl:eq:shz}).
A significantly better approach is not to use the same $S_{I,O}$ values for both stars, but to weight each star with its proper effective insolation constant instead, so that
\begin{equation}
\frac{L_A}{SA_{I}}\frac{1}{a^2_{I}}+\frac{L_B}{SB_{I}}\frac{1}{b^2_{I}} \le 1 \qquad \text{and} \qquad \frac{L_A}{SA_{O}}\frac{1}{a^2_{O}}+\frac{L_B}{SB_{O}}\frac{1}{b^2_{O}} \ge 1,
\label{eggl:eq:swhzbi}
\end{equation}
where $SA_{I,O} = S_{I,O}(T_{eff}(A))$ and $SB_{I,O} = S_{I,O}(T_{eff}(B))$ are the habitable flux limits for the inner and outer edges of the single star habitable zone using the effective temperatures of stars A and B, respectively \citep{eggl-2018}.
We are using the effective insolation limits for each individual star as weights to model how much star A affects the climate of a terrestrial planet compared to star B.
We shall, therefore, refer~to $SA_{I,O}$ and $SB_{I,O}$ as ``spectral weights''.
In a binary star system with two stars identical to our Sun, a planet would receive exactly one solar constant's worth of insolation with a solar spectral energy distribution ($SA=SB=1 s_\odot$) where
\begin{equation}
\frac{1 L_\odot\;\text{sr}}{a^2}+\frac{1 L_\odot\;\text{sr}}{b^2} = 1 s_\odot.
\label{eggl:eq:swhzbsimple}
\end{equation}
Hence, for $a=b$, the effective insolation from both stars at a star-planet distance of $a=\sqrt{2}$ au is equal to one solar constant.
% Both, the luminosity and the spectral weights are approximately constant for a given star, at least as long as the star is on the main sequence.
We can rewrite Equation (\ref{eggl:eq:swhzbi}) in a more concise form by introducing spectrally weighted luminosities
\begin{equation}
 \mathbb A_{I,O}=L_A/SA_{I,O}, \qquad \mathbb B_{I,O}=L_B/SB_{I,O},
 \label{eggl:eq:lsw}
\end{equation}
where $[\mathbb A]=[\mathbb B]=[\text{au}^2]$.
The following implicit expressions in the distances $a$ and $b$ describe
\begin{equation}
 \text{the inner border:} \;\; \frac{\mathbb A_{I}}{a^2_{I}}+\frac{\mathbb B_{I}}{b^2_{I}} = 1 \quad \text{and the outer border:}\;\; \frac{\mathbb A_{O}}{a^2_{O}}+\frac{\mathbb B_{O}}{b^2_{O}} = 1
\label{eggl:eq:swhz}
\end{equation}
of the isophote-based habitable zone in binary star systems. Inserting
\begin{equation}
a=[(x+d/2)^2+y^2 + z^2]^{1/2}, \qquad b=[(x-d/2)^2+y^2 +z^2]^{1/2},
\label{eggl:eq:rarb}
\end{equation}
in Equation (\ref{eggl:eq:swhz}), where x, y, and z are coordinates with respect to the origin of a convenient coordinate center
that has both stars along the x axis at a distance of $d$ from each other, we see that the isophote-based habitable zone borders depend on the mutual distance $d$ between the two stars.
Since $d$ changes over time for binary star systems on eccentric orbits, the isophote-based habitable zone also varies with time \citep{mueller-haghighipour-2014}.
For co-planar systems, i.e., $z=0$, one can find analytic solutions
to Equation (\ref{eggl:eq:swhz}) by expressing $y=F(x,d,\mathbb A, \mathbb B)$. This leads to a quartic equation
that, although unwieldy, can be solved \citet{cuntz-2014, cuntz-2015}. Alternatively,
one can use either numerical methods or analytic approximations
based on fixed-point iterations to solve Equation (\ref{eggl:eq:swhz}).
The latter approach will be discussed shortly.

\begin{table}
\caption{Data sheet for $\alpha$ Centauri and Kepler-35,
see \citet{welsh-2012,thevenin-2002,kervella-2003,zombeck-2006,quarles-2016}. The following orbital parameters have been used for $\alpha$ Centauri: $a_b=23.52$ au, $e_b=0.5179$, and Kepler-35: $a_b=0.17617$ au, $e_b=0.1421$.
SSHZ$_{I,O}$ symbolizes the inner and outer
single star HZ border, respectively. }
\label{eggl:tab:stars}       % Give a unique label
\begin{tabular}{p{2cm}p{1.5cm}p{1.5cm}p{1.5cm}p{1.5cm}p{1.5cm}p{1.5cm}}
\hline
 & $L $ & $T_{eff}$ & $R $ & $m$ & SSHZ$_I$  & SSHZ$_O$  \\
Star & [$L_\odot$] & [K] & [$R_\odot$] & [$M_\odot$] &[au] &[au]\\
\hline
 $\alpha$ Centauri A & 1.52 & 5790 & 1.227 & 1.1 &  1.17 &  2.06 \\
 $\alpha$ Centauri B & 0.50 & 5260 & 0.865 & 0.93 & 0.69 & 1.24 \\
 Kepler-35 A & 0.94 & 5606 & 1.03 & 0.89 & 0.93 & 1.65 \\
 Kepler-35 B &0.41 & 5202 & 0.79 & 0.81 & 0.63&  1.13         \\
% F5V & 2.5 & 6540 & 1.2 & 1.3 &1.44& 2.49\\
% G2V & 1   & 5777 & 1   & 1 & 0.95& 1.68\\
% M5V 	& 0.008 	& 	3120  & 0.32	&	 0.21 & 0.09& 0.18 \\
\hline
\end{tabular}
\end{table}

Figure \ref{eggl:fig:stype-iso} compares single star habitable zone insolation limits, where the contribution of the second star is ignored, to isophote-based habitable zones.
In close binary star systems with $\alpha$ Centauri-like stellar components there is a clear difference between single star habitable zones and isophote-based habitable zones.
The combined flux of the two stars causes the isophote-based habitable zones around both stars to extend toward each other.
In S-type systems the largest displacement of the isophotes is registered along the line connecting the
centers of the two stars. For binary stars on elliptic orbits, the isophote displacement is a function of their mutual distance $d$ and, thus, time.
Generally speaking, the stars influence each other's habitable zone most during their closest approach, i.e., near the pericenter $q_b=a_b(1-e_b)$,
where $a_b$ is the semi-major axis in [au] and $e_b$ the orbital eccentricity of the binary.
Circumstellar habitable zones will be least affected by the respective companion stars
when they are at apocenter $Q_b=a_b(1+e_b)$. For circular orbits, we~have $e_b=0$ and, thus, $Q_b=q_b=a_b=d$.

\begin{figure}[]
\centering
\includegraphics[scale=1]{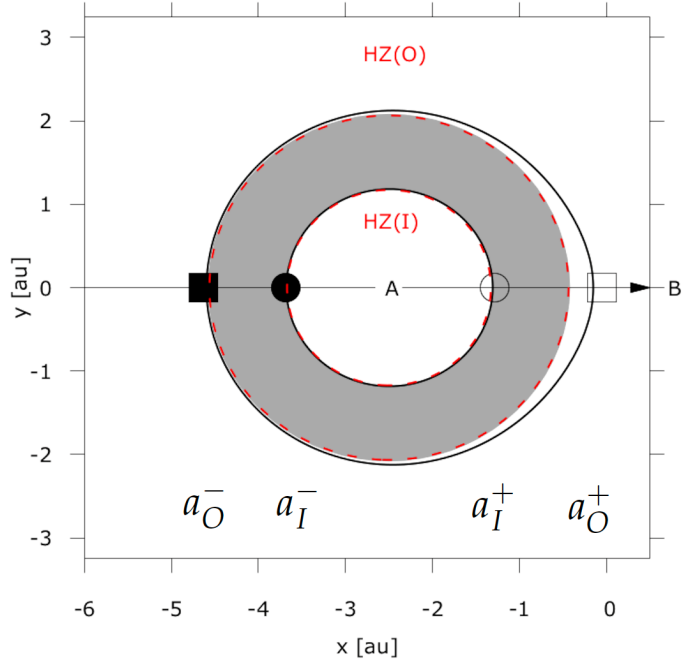}
\caption{A zoom on the brighter star in Figure \ref{eggl:fig:stype-iso}.
The distances $a^{(+,-)}_{I,O}$ calculated from Equations~(\ref{eggl:eq:rap}) and (\ref{eggl:eq:ram}) are intersections of the isophotes
corresponding to habitable zone limit insolation values with the line connecting the center of both stars (y = 0). The inner (HZ (I)) and outer (HZ (O)) borders of the
single star habitable zone are given by red dashed lines. The intersections of the isophotes with the abscissa
 for the inner HZ border $a^{(+,-)}_I$ are shown with empty and full circles, whereas the outer borders $a^{(+,-)}_O$ are denoted by
empty and full squares, respectively. The radiative habitable zone as defined by \citet{cuntz-2014} is the area shown in gray. The radiative habitable zone constitutes
a hollow sphere with inner radius $a_I^{+}$ and outer radius $a^{-}_O$.}
\label{eggl:fig:csrhz}      % Give a unique label
\end{figure}

To quantify the largest shift in isophote-based habitable zone borders we can substitute $b=d-a$ in
Equation (\ref{eggl:eq:swhz}). We shall only consider distances along line connecting the star centers (line of centers) for the moment. Multiplying both sides of the equation
by $a$ we can interpret the resulting expression as a fixed-point iteration,
\begin{equation}
 a_{j+1}= \frac{\mathbb A}{a_{j}}+\frac{\mathbb B\; a_{j}}{(d-a_{j})^2},
\label{eggl:eq:fixp1}
\end{equation}
where  $j$ is an integer and $a_{j+1}$ denotes the distance of the (spectrally weighted) isophotes centered around star A after the $j+1$ iteration step.
As starting points we use the classical habitable zone limits,
\begin{equation}
 a_{j=0}= \pm\sqrt{\mathbb A}.
\label{eggl:eq:fixp2}
\end{equation}
Choosing the positive roots of Equation (\ref{eggl:eq:fixp2}) and stopping after the first iteration, we find the following approximation for the new isophote positions
between the two stars
\begin{equation}
 a^+\approx \sqrt{\mathbb A}\left(1+\frac{\mathbb B}{(d-\sqrt{\mathbb A})^2}\right).
\label{eggl:eq:rap}
\end{equation}
Inserting the respective spectral weights, i.e. ${\mathbb A}_I$, ${\mathbb B}_I$ or ${\mathbb A}_O$, ${\mathbb B}_O$ into Equation (\ref{eggl:eq:rap}) the resulting values $a^+_{I,O}$ represent the new inner and outer habitable zone borders around star A in the direction of star B.
We can use the negative square root of Equation (\ref{eggl:eq:fixp2}) to calculate the habitable zone borders on the opposite side of star A.
\begin{equation}
 a^-\approx -\sqrt{\mathbb A}\left(1+\frac{\mathbb B}{(d+\sqrt{\mathbb A})^2}\right).
\label{eggl:eq:ram}
\end{equation}
Figure \ref{eggl:fig:csrhz} illustrates the above points showing a zoom on star A in Figure \ref{eggl:fig:stype-iso}.
We see that the distances $a^{+,-}_{I,O}$ are intersections of the isophotes corresponding to the habitable zone limits with the line of centers.
This means that the isophote-based habitable zone borders along the line connecting the two stars are given by
\begin{eqnarray}
IHZ_A^+&=&a^{+}\approx \sqrt{\mathbb A}\left(1+\frac{\mathbb B}{(d-\sqrt{\mathbb A})^2}\right), \label{eggl:eq:csihza} \\
IHZ_A^-&=&a^{-}\approx -\sqrt{\mathbb A}\left(1+\frac{\mathbb B}{(d+\sqrt{\mathbb A})^2}\right),
\label{eggl:eq:csihz}
\end{eqnarray}
in the direction of the second star B (+), and in the opposite direction ($-$), respectively. Equation (\ref{eggl:eq:csihza}) has two results depending on whether ${\mathbb A}_{I}$ and ${\mathbb B}_{I}$ or ${\mathbb A}_{O}$ and ${\mathbb B}_{O}$ are chosen. The same holds for Equation (\ref{eggl:eq:csihz}). The four points given by Equations (\ref{eggl:eq:csihza}) and (\ref{eggl:eq:csihz}) characterize the extent of the isophote-based habitable zone.
The dimension of isophote-based habitable zone borders is that of a length, in our case [au], as can be seen from Equations (\ref{eggl:eq:swhz}), (\ref{eggl:eq:rap}) and (\ref{eggl:eq:ram}).

The difference between the single star habitable zone and the corresponding isophotes
is largest in the direction of star B, e.g., $|IHZ^+_{A,\, O}| > |IHZ^-_{A,\, O}|$.
The asymmetry between the two directions~(+,$-$) leads to a deformation of the single star habitable zone into the tear-shaped isophote-based habitable zones.
It is evident from Figure \ref{eggl:fig:stype-iso} that in some cases the single star habitable zone provides a reasonable enough approximation to the isophote-based habitable zone. As a rule of thumb that is the case for two sun-like stars that have a pericenter distance of no less than 10$\;$au.
To determine where the single star habitable zone is no longer a good proxy for the actual isophote-based habitable zone in S-type configurations
we can use Equations (\ref{eggl:eq:fixp1}) and (\ref{eggl:eq:fixp2}) to calculate the relative displacement $\Delta a$ of isophote-based habitable zone borders with respect to the single star habitable zone borders. This yields
\begin{equation}
     \Delta a=\frac{a^1-a^0}{a^0}=\frac{\mathbb B}{(d-\sqrt{\mathbb A})^2}.
\label{eggl:eq:deltar}
\end{equation}
%As the radiative contribution of each star is proportional to $r^{-2}$, the largest displacement is expected for the outer edges of the HZ in between the two stars.

\begin{figure}[]
\centering
%\begin{tabular}{ll}
\includegraphics[scale=1.1]{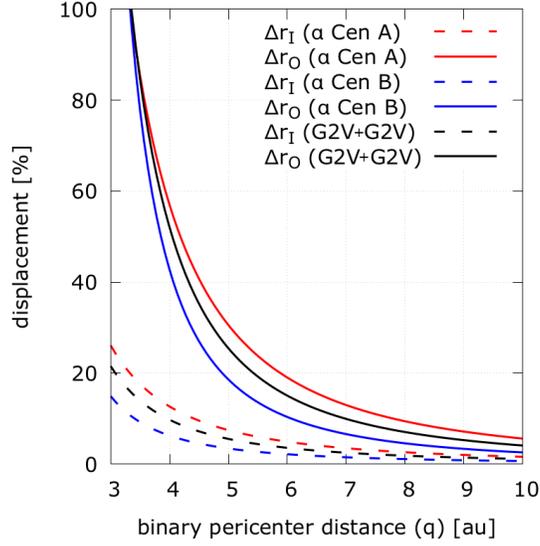}
%\end{tabular}
\caption{A visualization of Equation (\ref{eggl:eq:deltar}) showing the maximum displacement of the inner ($I$, dashed) and outer ($O$,~continous) border
of the isophote-based habitable zone as a function of the binary orbit pericenter distance $q$.
The displacement is given in relative to the original single star habitable zone borders. A $\Delta a=100\%$ means that the new border is twice as far
from its host star than the single star habitable zone pendant. We consider $\alpha$ Centauri-like systems and a binary consisting of two sun-like (G2V) stars.}
\label{eggl:fig:deltar}       % Give a unique label
\end{figure}

Figure \ref{eggl:fig:deltar} illustrates the displacement of the isophote-based habitable zone borders as a function of the binary star's pericenter distance ($q$).
Three cases are shown, one for $\alpha$ Cen A, one for $\alpha$~Cen B and one for a system of solar twins.
One can see that the displacement is always larger for the outer isophote-based habitable zone borders than for the inner ones. Absent dynamical constraints this means that the presence of the second star leads to a
net growth of the isophote-based habitable zone compared to the single star habitable zone.
For very close binaries, on the other hand, the individual isophote-based habitable zones of both stars merge into a single circumbinary isophote-based habitable zone as shown in Figure \ref{eggl:fig:ptype-iso}.
In~order to calculate circumbinary isophote-based habitable zone borders we insert Equation (\ref{eggl:eq:rarb}) into Equation~(\ref{eggl:eq:swhz}) and
define $\delta:=d/2$.
Still assuming a coplanar configuration ($z=0$) this yields
\begin{equation}
\frac{\mathbb A}{(x+\delta)^2 + y^2}+\frac{\mathbb B}{(x-\delta)^2 + y^2} = 1
\label{eggl:eq:ptxy}
\end{equation}
For very small separations of the binary $\delta\rightarrow 0$, Equation (\ref{eggl:eq:ptxy}) gives
\begin{equation}
 c_{I,O}=\sqrt{\mathbb A_{I,O}+\mathbb B_{I,O}},
\label{eggl:eq:pthzb0}
\end{equation}
where  $c=\sqrt{x^2+y^2}$ is the distance of the planet to the origin of the coordinate system.
For compact, equal mass binary stars $c$ is the distance to the barycenter of the two stars.
A crude approximation to the isophote-based habitable zone in such systems can be constructed if we assume that both stars have exactly the same location at the origin of our coordinate system.
The circumbinary isophote-based habitable zone then resembles a classical habitable zone around a 'hybrid-star' featuring the combined spectrally weighted luminosities of both stars. The inner and outer borders of the circumbinary isophote-based habitable zone are then approximated by
\begin{eqnarray}
IHZ_{AB,\, I}& \approx & \sqrt{\mathbb A_{I}+\mathbb B_{I}}\;,\\
IHZ_{AB,\, O}& \approx & \sqrt{\mathbb A_{O}+\mathbb B_{O}}\;.
\label{eggl:eq:cbihzapprox}
\end{eqnarray}
This approximation is only reasonable for systems where $d\ll c_I$ as Figure \ref{eggl:fig:ptype-iso} reveals.
If the distance between the stars is comparable to the inner limit of the 'hybrid-habitable zone', then isophote-based habitable zone and radiative habitable zone limits can no longer be calculated using Equation (\ref{eggl:eq:pthzb0}).
Since the assumption that the single star habitable zone is a good starting point for the fixed-point iteration no longer holds
we have to construct another fixed-point iteration with $c_{I,O}=\sqrt{\mathbb A_{I,O}+\mathbb B_{I,O}}$ as initial guess.
Restricting Equation (\ref{eggl:eq:ptxy}) to the line of centers between the binary stars ($y=0$) we find
\begin{eqnarray}
 IHZ_{AB}^+ = x_{AB}^- &\approx & -\left(\mathbb A\frac{c+\delta}{c-\delta}+\mathbb B\frac{ c-\delta}{c+\delta}-\delta^2\right)^{1/2},\\
 IHZ_{AB}^- = x_{AB}^+ & \approx & \left(\mathbb A\frac{c-\delta}{c+\delta}+ \mathbb B\frac{c+\delta}{c-\delta}-\delta^2\right)^{1/2}
\label{eggl:eq:cbihz}
\end{eqnarray}
as approximants for the circumbinary isophote-based habitable zone borders. Once more, Equation~(\ref{eggl:eq:cbihz}) represents four equations as the corresponding spectral weights need to be accounted for in $\mathbb A_{I,O}$ and $\mathbb B_{I,O}$.

%Comparing the black and the purple lines in Figure \ref{eggl:fig:stype-iso} one can see that the isophote-based habitable zone of the second star is only
%mildly truncated by orbital instability. In contrast, the potentially habitable region around the primary is severely diminished by the presence
%of the second star.

\begin{figure}[]
\centering
%\begin{tabular}{ll}
\includegraphics[scale=1.1]{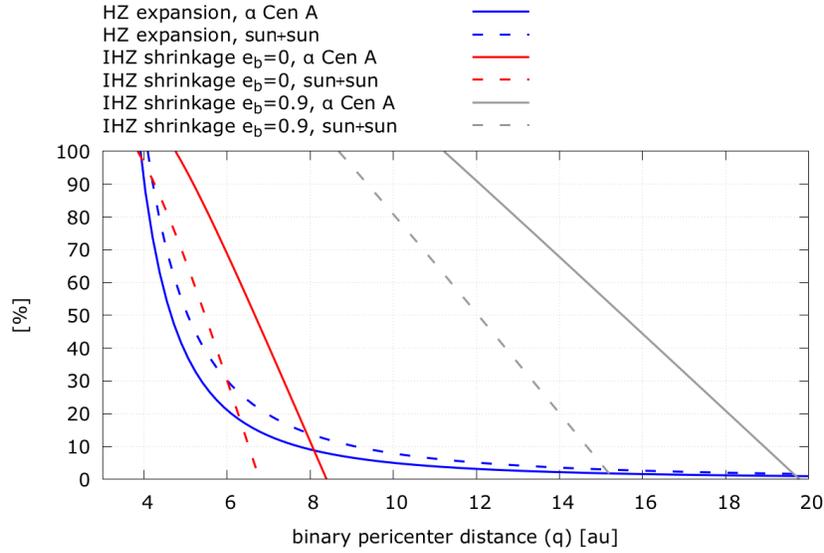}
%\end{tabular}
\caption{
The different rates of displacement of the isophote-based habitable zone borders in Figure~\ref{eggl:fig:deltar} lead to an expansion of the isophote-based habitable zone compared to the single star habitable zone.
At the same time orbital stability restrictions shrink the size of the isophote-based habitable zone. A~shrinkage of $50\%$ means that
the outer half of the isophote-based habitable zone is dynamically unstable whereas a shrinkage of $100\%$ means that no planet can
remain on a stable orbit in the isophote-based habitable zone. Expansion and shrinkage of the isophote-based habitable zone are computed with respect to the primary of an $\alpha$ Centauri like system and a
binary consisting of two sun-like G2V~stars.}
\label{eggl:fig:sx}       % Give a unique label
\end{figure}

%Which process is more important, the expansion of the isophote-based habitable zone due to the second source of radiation or the shrinkage of the isophote-based habitable zone due to orbital instability?
Figure \ref{eggl:fig:sx} shows the evolution of the isophote-based habitable zone size as a function of the binary star pericenter distance and orbital eccentricity.
For two sun-like stars on circular orbits, the isophote-based habitable zone borders can expand up to $\approx$25\% before orbital instability starts to chew away on the outer isophote-based habitable zone limit.
No stable circumstellar orbits are left in the isophote-based habitable zone when the sun-like binary stars orbit each other at a distance closer than 4 au.
Systems similar to $\alpha$ Centauri allow for the isophote-based habitable zone to grow around the primary for as little as $10\%$ before orbital instability sets in.
For binaries on highly eccentric orbits the isophote-based habitable zone is substantially truncated even at large pericenter distances. Such an example can be found in \citet{georgakarakos-2019}.

We conclude that if orbital stability of the planet is required, the maximum radiative contribution of the second star to the extent and location of the isophote-based habitable zone is relatively small.
Which systems are expected to have a circumstellar habitable zone that is not truncated due to orbital instability?
Combining Equation (\ref{eggl:eq:rap}) with a simplified form of Equation (\ref{eggl:eq:stabs}) we find that the entire circumstellar habitable zone is stable if the binary pericenter distance is larger than
\begin{equation}
q_b > 2 \mathbb A_O^{1/2} \left(1+\mathbb A_O^{1/3}\mathbb B_O^{1/3}\right),
\label{eggl:eq:cshzlim1}
\end{equation}
where $q_b=a_b(1-e_b)$ represents the pericenter distance of the stellar binary and $\mathbb{A}_O$ is the spectrally weighted luminosity for the outer edge of the single star habitable zone.
Choosing values for $\mathbb A_O$ and $\mathbb B_O$ of the actual $\alpha$ Centauri system Equation (\ref{eggl:eq:cshzlim1}) predicts a the stability of the entire circumstellar habitable zone around $\alpha$ Centauri A for $q_b>11.8\;$au. This is in good agreement with the results presented in \citet{quarles-2016}, that find the orbital stability limit around $\alpha$ Centauri A to be close to $2\;$au which is also the outer limit of the single star habitable zone \citep{kopparapu-2014}. $\alpha$ Centauri has a pericenter distance of $q_b\approx11.3\;$au.
Conversely, Equation (\ref{eggl:eq:cshzlim1}) suggests that the circumstellar habitable zone around $\alpha$~Centauri B would be stable even if $q_b\approx7.1\;$au. The more generous stability limit is due to the fact that $\alpha$ Centauri B is only half as luminous as our Sun. The outer border of the corresponding single star habitable zone is, therefore, at $1.24\;$au well within the dynamically stable region \citep{quarles-2016}.

For circumbinary planets, Equation (\ref{eggl:eq:stabp}) dictates that the distance between the planet and the center of mass of the binary must be several times
the semi-major axis of the stellar orbit in order to allow for stable configurations.
Using Equations (\ref{eggl:eq:stabp}) and (\ref{eggl:eq:pthzb0})
we can define an approximate limit condition for the existence of dynamically stable circumbinary isophote-based habitable zones, namely
\begin{equation}
Q_b^2<\frac{\mathbb{A}_I+\mathbb{B}_I}{6},
\label{eggl:eq:cbhzlim}
\end{equation}
where $Q_b=a_b(1+e_b)$ is the apocenter of the binary star orbit, and $\mathbb{A}_I$ and $\mathbb{B}_I$ are the spectrally
weighted luminosities for the inner edge of the respective single star habitable zones, see
Equation (\ref{eggl:eq:lsw}).
If the binary star orbit is tight enough, circumbinary isophote-based habitable zones are dynamically stable.
As we shall explore in the next sections, dynamical stability and isophote-based habitable zone constraints may not be enough, however, to determine where Earth-like planets can be habitable
in binary star systems.

Moreover, Equations (\ref{eggl:eq:rap}) and (\ref{eggl:eq:ram}) show that the deformation of the single star habitable zone is a function of the binary star distance $d$.
The latter is, however, time dependent for all but systems where the stars have a circular orbit with respect to their common center of mass.
To cope with this issue, \citet{mueller-haghighipour-2014} introduced
rotating, pulsating isophote-based habitable zones. Those can be thought of as analogous to pulsating coordinate systems or zero velocity curves that are sometimes used to study the elliptic restricted three body problem \citep{szebehely1969theory}. Having time-varying habitable zone borders means that isophote-based habitable zones sweep over planets on relatively short timescales.
This leads to the problem of determining to which degree planets that are only partly inside habitable zones are actually habitable - a
 topic of an ongoing investigation \citep{williams-pollard-2002,dressing-2010,bolmont-2016}.

% ##################################
\section{Radiative Habitable Zones}
\label{eggl:sec:rhz}

%whereas Radiative Habitable
%Zones (Cuntz 2014, 2015) provide habitable zone limits
%for planets on circular circumstellar and circumbinary
%orbits.

\begin{figure}[H]
\centering
%\begin{tabular}{ll}
\includegraphics[scale=0.35]{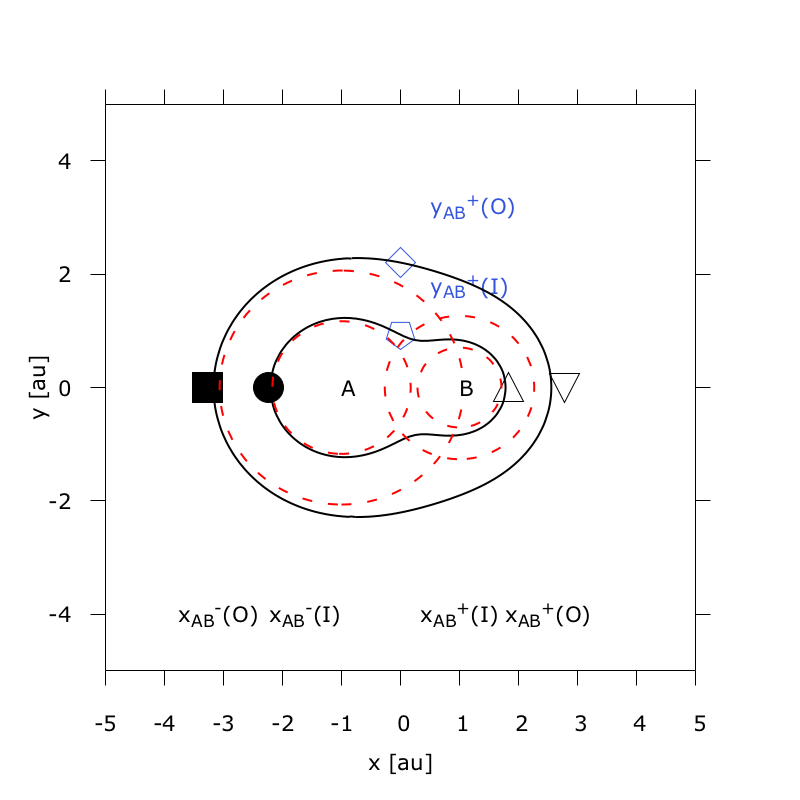}
\includegraphics[scale=0.35]{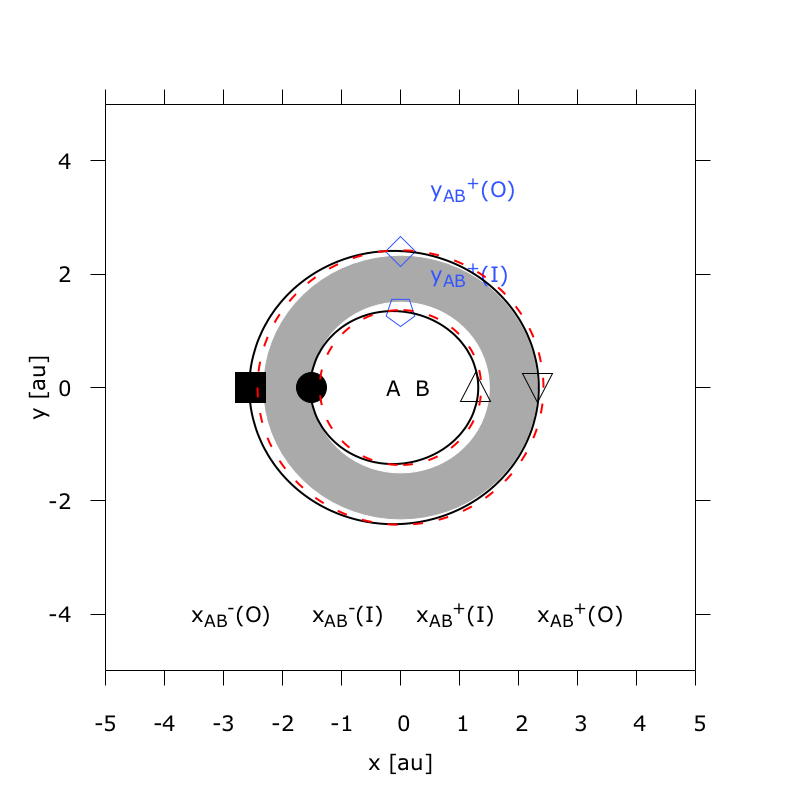}
%\end{tabular}
\caption{Isophote-based habitable zones and radiative habitable zones for close binary star configurations (P-type). The graphs are for configurations similar to $\alpha$ Centauri only on circular orbits with distances
$d=3$ au (left panel) and $d=0.5$ au (right panel).
The inner and outer borders of the single star habitable zone are given by red dashed lines. The red dashed lines in the right panel would represent the habitable zone, if both stars were located at the origin of the graph.
The isophote-abscissa intersection points in the left panel are derived from Equations (\ref{eggl:eq:rap}) and (\ref{eggl:eq:ram}),
whereas the ones on the right result from Equation (\ref{eggl:eq:cbihz}), respectively.
The radiative habitable zone vanishes for the system in the left panel. The closer system on the right has a radiative habitable zone shown in gray.}
\label{eggl:fig:cbrhz}      % Give a unique label
\end{figure}

One of the issues of using isophote-based habitable zones is that planetary orbits in binary star systems do not follow isophotes.
As a consequence \citet{cuntz-2014} introduced the so-called ``radiative habitable zone'' (RHZ). In analogy to single star habitable zones the radiative habitable zone is based on the assumption that planets are moving on circular orbits either
around one or both of the stars forming the binary. The radiative habitable zone is then defined as the largest spherical shell
that can be inscribed in the isophote-based habitable zone. As such its limits can be approximated through:
\begin{eqnarray}
 RHZ_{A,\;I}&=&|IHZ_{A,\;I}^+|\approx \sqrt{\mathbb A_I}\left(1+\frac{\mathbb B_I}{(d-\sqrt{\mathbb A_I})^2}\right), \\
 RHZ_{A,\;O}&=&|IHZ_{A,\;O}^-|\approx \sqrt{\mathbb A_O}\left(1+\frac{\mathbb B_O}{(d+\sqrt{\mathbb A_O})^2}\right).
\label{eggl:eq:csrhz}
\end{eqnarray}
Due to its symmetry and in contrast to Equation (\ref{eggl:eq:cbihz}) the radiative habitable zone has only one equation for its inner and one equation for its outer edge.
For the exact expressions we would like to refer the reader to
%\citeauthor{cuntz-2014} (\citeyear{cuntz-2014}).
\citet{cuntz-2014, cuntz-2015}.
Figure \ref{eggl:fig:csrhz} shows the radiative habitable zone around star A of an $\alpha$ Centauri-like system with $d=5$ au. The radiative habitable zone is smaller than the isophote-based habitable zone, but both start to coincide for large distances between the primary and secondary star. In fact, in the limit $d\rightarrow\infty$, the radiative habitable zone, isophote-based habitable zone and single star habitable zone are identical.
Similar to the circumstellar (S-type) case, we can define a circumbinary (P-type) radiative habitable zone as
\begin{eqnarray}
 RHZ_{AB, \;I}&=&|IHZ_{AB, \;I}^-|\approx \left(\mathbb A_I\frac{\sqrt{\mathbb A_I + \mathbb B_I}+\delta}{\sqrt{\mathbb A_I + \mathbb B_I}-\delta}+ \mathbb B_I\frac{\sqrt{\mathbb A_I + \mathbb B_I}-\delta}{\sqrt{\mathbb A_I + \mathbb B_I}+\delta}-\delta^2\right)^{1/2} , \\
 RHZ_{AB, \;O}&=&|IHZ_{AB, \;O}^+|\approx \left(\mathbb A_O\frac{\sqrt{\mathbb A_O + \mathbb B_O}-\delta}{\sqrt{\mathbb A_O + \mathbb B_O}+\delta}+\mathbb B_O\frac{ \sqrt{\mathbb A_O + \mathbb B_O}+\delta}{\sqrt{\mathbb A_O + \mathbb B_O}-\delta}-\delta^2\right)^{1/2}.
\label{eggl:eq:cbrhz}
\end{eqnarray}
Contrary to circumstellar radiative habitable zones, the existence of circumbinary radiative habitable zones is not guaranteed. For certain stellar distances $d$ the circumbinary isophote-based habitable zone starts to deform and finally separate into individual stellar isophote-based habitable zones.
Thus, at~a certain limit distance $d$ the shape of the isophote-based habitable zone does not permit
to inscribe spherical shells anymore (see Figure \ref{eggl:fig:cbrhz}).
To determine when that is the case, let us investigate the isophote intersections with the ordinate of our coordinate system.
Letting $x=0$ in Equation (\ref{eggl:eq:ptxy}) we find that
\begin{equation}
 y_{AB}=\pm\sqrt{\mathbb A+\mathbb B-\delta^2}.
\label{eggl:eq:yab}
\end{equation}
Two important constraints follow from the above expression,
one for the existence of the radiative habitable zone
\begin{equation}
 y_{AB}^+(O)\geq x_{AB}^+(I),
\label{eggl:eq:exrhz}
\end{equation}
and the other for the merging of the individual isophote-based habitable zones of star A and B into a single circumbinary $IHZ_{AB}$
\begin{equation}
\mathbb A_{I,O}+\mathbb B_{I,O} \geq  d^2/4.
\label{eggl:eq:12ihz}
\end{equation}
Equation (\ref{eggl:eq:exrhz}) tells us that we can only inscribe spherical shells, if the inner radiative habitable zone border does not intersect the outer isophote-based habitable zone border.
On the other hand, Equation~(\ref{eggl:eq:12ihz}) predicts that isophote-based habitable zones merge whenever the distance between the stars is smaller than twice the square root of the sum of the spectrally weighted luminosities.

%%%%%%%%%%%%%%%%%%%%%%%%%%%%%%%%%%%%%%%%%%
\section {Dynamically Informed Habitable Zones}
\label{eggl:sec:dhz}
\begin{figure}[]
\centering
\includegraphics[scale=.22]{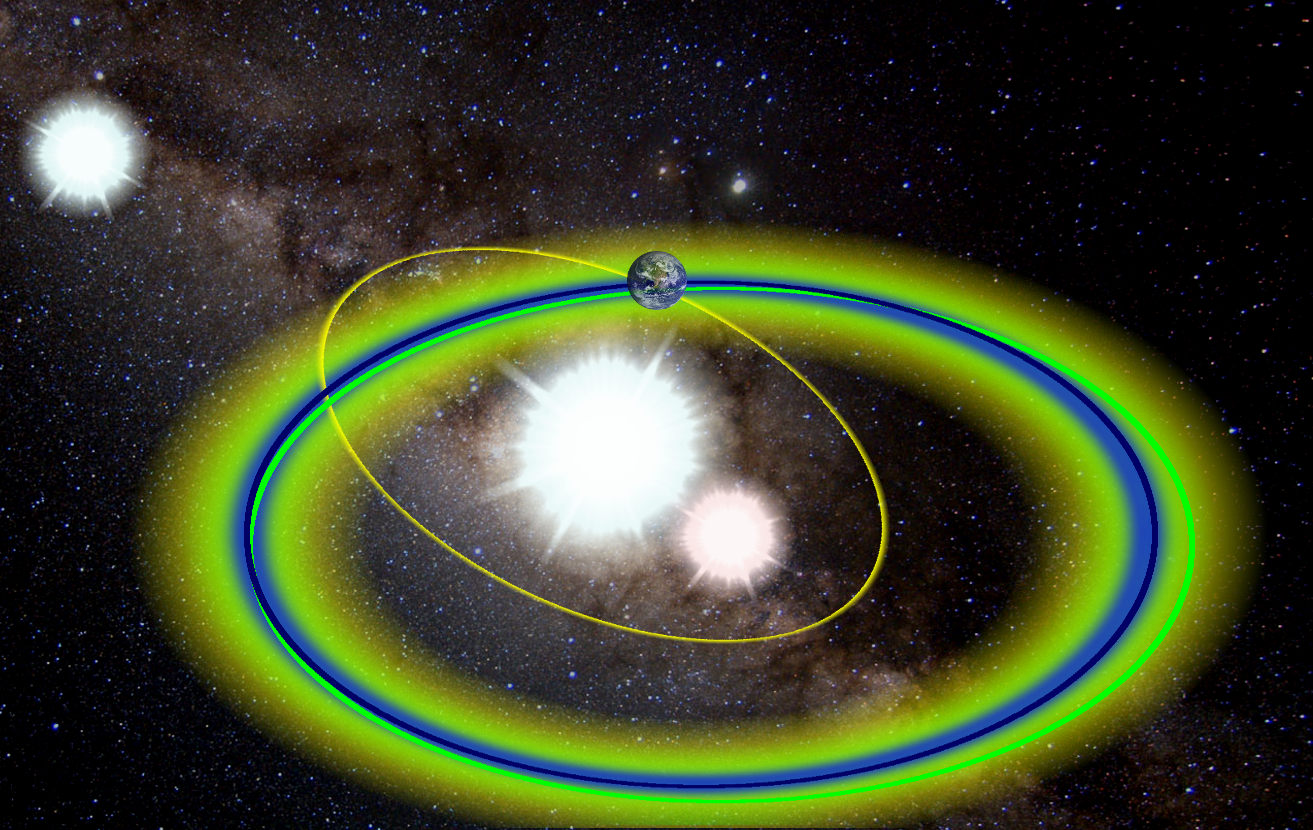}
\caption{Dynamically informed habitable zones. The permanently habitable zone in blue represents the most conservative limits. A planet started in the permanently habitable zone will never exceed habitable insolation limits throughout its orbital evolution. In contrast, a planet in the averaged habitable zone (yellow) can experience ample excursions beyond habitable insolation limits as long as the insolation average permits liquid water near the surface of the planet. The green curve represents an orbit of the extended habitable zone as defined in \citet{eggl-et-al-2012}.}
\label{eggl:fig:dhz}       % Give a unique label
\end{figure}

Radiative habitable zones are largely sufficient to provide an idea as to where planets can be located in a binary star system and still retain liquid water near their surface. If we take into account that planets move on
perturbed Keplerian orbits we have to acknowledge the fact that the amount of light a planet receives from the binary star can change drastically with time \citep{eggl-et-al-2012, eggl-2018}.
How planetary atmospheres react to such changes is a matter of ongoing investigation
\citep{williams-pollard-2002, spiegel-et-al-2010,dressing-2010, forgan-2016, popp-eggl-2017,moorman-2019}.
To reduce the complexity of the problem  \citet{eggl-et-al-2012} introduced the concept
of climate inertia on planetary habitability.
An analogy would be the apparent thermal inertia~\citep{price-1985} for celestial objects, such as the Moon \citep{williams-2017} and the Earth~\citep{ermida-2019}.
If the climate of a planet reacts to changes
in insolation with very little latency (low climate inertia), then the planet must remain inside the so-called permanently habitable zone (PHZ) to allow for liquid water to exist near its surface.
For a planet to reside in the permanently habitable zone, maximum and minimum values of the insolation function must not exceed habitable limits at any time.
The permanently habitable zone corresponds to a strict, ``classical'' definition of a habitable zone when orbital dynamics are taken into account.
A more relaxed definition allows some parts of the planetary orbit to lie outside the permanently habitable zone.
Following the argument of \citet{williams-pollard-2002} that the atmosphere and oceans of a planet can buffer
insolation variability we can define an averaged habitable zone (AHZ).
Insolation extrema are ignored as long as the time averaged insolation stays within habitable bounds.
Formally the above habitable zones are defined as
\begin{eqnarray*}
PHZ: & \max (\mathbb S_I) \leq 1\qquad &\text{and} \qquad  \min (\mathbb S_O) \geq 1 \\
%EHZ: & \left \langle \mathbb I_I \right \rangle_t +\sigma_I \leq 1 \; \text{and} \;
%\left \langle \mathbb I_O \right \rangle_t -\sigma_O \geq 1 \\
AHZ: &\left \langle \mathbb S_I \right \rangle \leq 1  \qquad &\text{and} \qquad  \langle  \mathbb S_O \rangle \geq 1,
\end{eqnarray*}
where
\begin{equation}
\mathbb S_{I,O}(t) = \frac{L_A}{SA_{I,O}}a^{-2}(t)+\frac{L_B}{SB_{I,O}}b^{-2}(t)
\end{equation}
is the combined spectrally weighted insolation on the planet, a function of time, and $\left \langle \mathbb S \right \rangle$ denotes the time-averaged combined stellar insolation. Again, $a$ and $b$ are the distances between the planet and star A and the planet and star B, respectively, and subscripts $I,O$ represent the inner and outer edge of the habitable zone.
%and $\sigma^2$ its variance.
%The EHZ always falls between the PHZ and AHZ borders so that we shall focus on deriving PHZ and AHZ borders in the following. We refer the reader to
% \cite{eggl-et-al-2012} for more details on calculating EHZs.
Note that all dynamically informed habitable zones neither depend on angular variables nor on time.
Consequently, permanent habitable zone and averaged habitable zone form concentric rings around the center of reference, just as the classical habitable zone, see Figure \ref{eggl:fig:dhz}.

\subsection{Circumstellar Habitable Zones}
\begin{figure}[H]
\centering
%\begin{tabular}{ll}
\includegraphics[scale=.58]{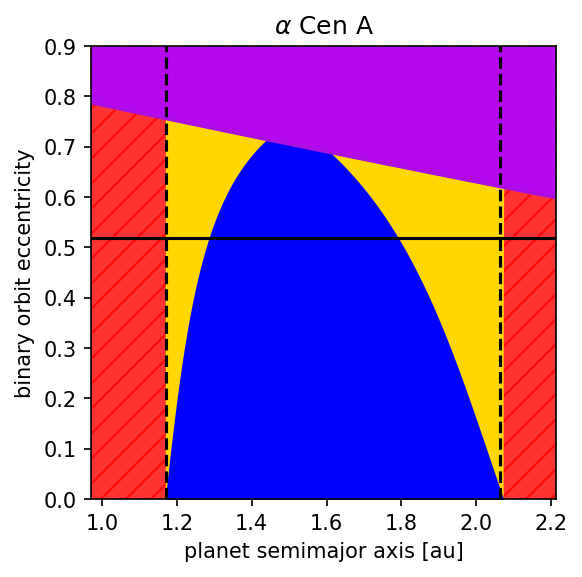}
\includegraphics[scale=.58]{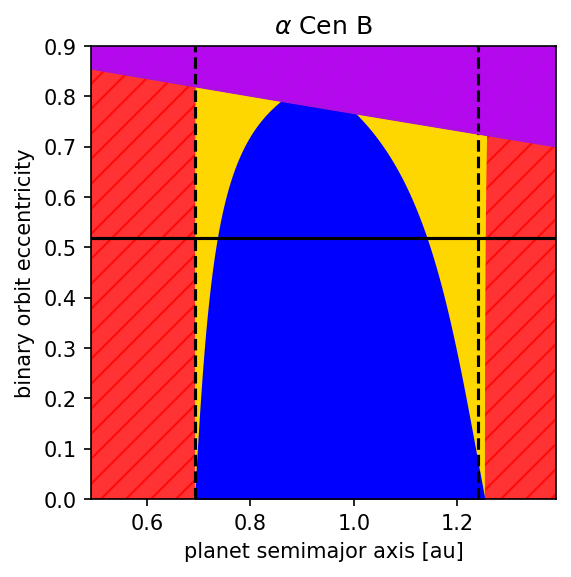}
\includegraphics[scale=.58]{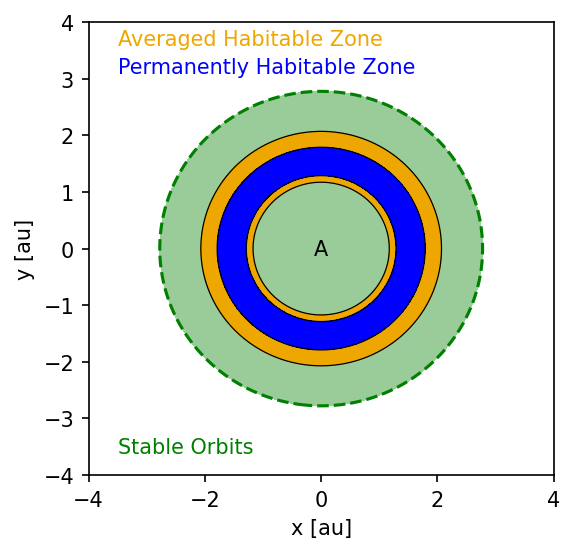}
\includegraphics[scale=.58]{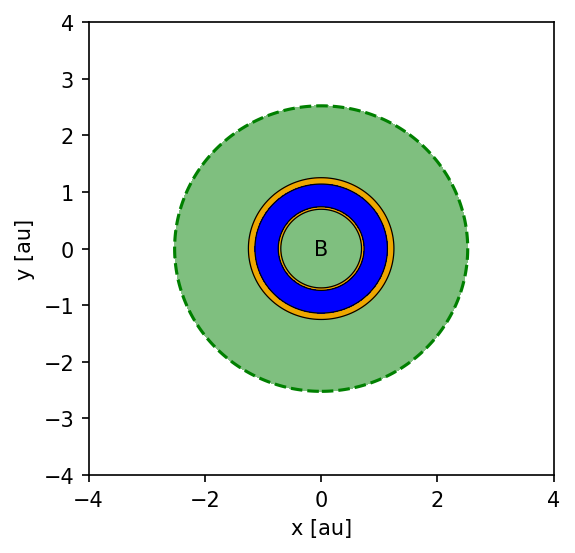}
%\end{tabular}
\caption{Dynamically informed habitable zones around $\alpha$ Centauri A and B. The top panels show how the extent of dynamically informed habitable zones changes with the orbital eccentricity of the system. Permanently habitable zones (blue), averaged habitable zones (yellow), dynamically unstable zones (purple) and non-habitable regions (red) are presented. The vertical, dashed lines describe single star habitable zone limits. The horizontal line represents the actual $\alpha$ Centauri system. The bottom panels represent a top down view on the actual system. Dynamically stable circumstellar zones around star A and B are colored green. }
\label{eggl:fig:dihz-s}  % Give a unique label
\end{figure}

Permanently habitable zones in S-type systems are regions where the following conditions hold:
\begin{eqnarray}
PHZ_I&:\quad \frac{\mathbb A_I} {q_p^2 }+\frac{\mathbb B_I}{(q_p-q_b)^2}\leq 1, \label{eggl:eq:phz0} \\
PHZ_O&:\quad \frac{\mathbb A_O} {Q_p^2}+\frac{\mathbb B_O}{(Q_p-Q_b)^2}\geq 1, \nonumber
\end{eqnarray}
with $q_p=a_p(1-e_p^{max})$, $q_b=a_b(1-e_b)$, $Q_p=a_p (1+e_p^{max})$ and $Q_b=a_b(1+e_b)$.
The maximum eccentricity the planetary orbit attains during its evolution is called $e_p^{max}=\max_t (e_p(t))$. The maximum
eccentricity $e_p^{max}$ of the planet depends on the orbital elements of the planet and the binary in a non-trivial way (\citet{eggl-et-al-2012}, see the Appendix therein). Equation (\ref{eggl:eq:phz0}) can be solved numerically with respect to $a_p$, however. The two solutions then correspond to the inner and outer permanently habitable zone limits.  If the initial orbit of a potentially habitable world supports a semi-major axis in the range $PHZ_I\leq a_p\leq PHZ_O$, orbital evolution will never carry the planet beyond habitable insolation limits.
In systems where the binary does not influence the orbit of the planet strongly, i.e., $e_p^{max} \ll 1$, we can estimate the permanent habitable zone borders analytically via
\begin{eqnarray}
PHZ_I&\approx \frac{\mathbb A_I} {q_p}+\frac{\mathbb B_I\,  q_p}{(q_p-q_b)^2}, \label{eggl:eq:phz2} \\
PHZ_O&\approx \frac{\mathbb A_O} {Q_p}+\frac{\mathbb B_O\, Q_p}{(Q_p-Q_b)^2},
\end{eqnarray}
where we can use $a_{p}=(\mathbb A_{I,O})^{1/2}$ as initial guesses to calculate $q_p$ and $Q_p$, respectively.
When formulating the maximum insolation condition for the inner edge of the permanent habitable zone we have assumed that the radiative contribution of star B does not overpower
that of star A. This condition reads
\begin{equation}
\mathbb B_I<(q_b-\sqrt{\mathbb A_I})^2
\end{equation}
If the planet receives more light at its apocenter than at its pericenter and Equation (\ref{eggl:eq:phz2}) must be adapted accordingly and
\begin{equation}
PHZ_I\approx \frac{\mathbb A_I} {Q_p}+\frac{  \mathbb B_I\,Q_p}{(Q_p-q_b)^2}. \label{eggl:eq:phzi}
\end{equation}
%valid if $B(I)<4 A(I)$ more precisely $B<A+2A^(4/3)B^(1/3)$.
If the climate of the planet has a high capacity to buffer changes in insolation, averaged habitable zone limits can be derived from insolation averages.
To simplify the calculation of planetary insolation averages \citep{eggl-et-al-2012}
have introduced the so-called equivalent radii. Equivalent radii are constant distances with respect to the host star that yield the same average amount of insolation a planet would receive, were it on an elliptic orbit.
In other words, the equivalent radius of a star-planet system corresponds to the semi-major axes of a circular orbit where a planet receives the same average insolation as it would on the original elliptic orbit.
In contrast to \citet{eggl-et-al-2012} the equivalent radii $\bar r_p$ and $\bar r_b$ are here chosen so as to be consistent with two body insolation averages. In other words,
\begin{equation}
\langle \mathbb S_A \rangle = \frac{1}{P}\int_0^P \frac{\mathbb A}{r_p^2(t)}dt \approx  \frac{\mathbb A}{a_p^2(1-\langle e_p^2\rangle)^{1/2}} :=\frac{\mathbb A}{\bar r_p^2}.
\end{equation}
where $r_p(t)=a_p(1-e_p^2)/(1+e_p\cos f_p(t))$, $\bar r_p=a_p (1-\langle e_p^2 \rangle)^{1/4}$, $f_p(t)$ is the true anomaly of the planet, $P$ is the orbital period of the planet and $\langle e_p^2\rangle$ is the averaged over time squared planetary eccentricity.
The equivalent radius for the secondary star with respect to the host star is \mbox{$\bar r_b=a_b(1-e_b^2)^{1/4}$}.
Averaging over the binary and the planetary orbit we find the conditions for the inner and outer border of the averaged habitable zone:
\begin{eqnarray}
AHZ_I&:\quad \frac{\mathbb A_I} {\bar r_p^2}+\frac{\mathbb B_I}{\bar r_b^2-\bar r_p^2}\leq 1 \label{eggl:eq:phz1}\\
AHZ_O&:\quad \frac{\mathbb A_O} {\bar r_p^2}+\frac{\mathbb B_O}{\bar r_b^2-\bar r_p^2}\geq 1. \nonumber
\end{eqnarray}
Note, that the above equations can be solved numerically for $a_p$ to find precise values for the averaged habitable zone borders.
As the average squared eccentricity $\langle e_p^2 \rangle$ is very small in most cases, the~following analytic estimate provides a good approximation
% \begin{eqnarray}
% AHZ(I,O)&\approx \frac{\mathbb A_{I,O}} {\bar r_p (1-\langle e^2 \rangle)^{1/4}}+\frac{\mathbb B_{I,O} a_p}{\bar r_b^2-\bar r_p^2},
% \end{eqnarray}
% \begin{eqnarray}
% AHZ(I,O)&\approx \frac{\mathbb A_{I,O}} {\bar r_p}+\frac{\mathbb B_{I,O}\bar r_p}{\bar r_b^2-\bar r_p^2},
% \end{eqnarray}
\begin{eqnarray}
AHZ_{I,O}&\approx \sqrt{\mathbb A_{I,O}}\left(1+\frac{\mathbb B_{I,O}}{a_b^2\sqrt{1-e_b^2}-\mathbb A_{I,O}}\right)
\end{eqnarray}
where we used $\bar r_p\approx(\mathbb A_{I,O})^{1/2}$.
Secular orbit evolution theory yields relatively compact expressions for $e_p^{max}$ and $\langle e^2_p\rangle$, see, for instance, \citet{andrade-2017}.
For planets on initially circular orbits those expressions read
\begin{equation}
e_p^{max}=2\epsilon, \qquad \langle e^2_p\rangle= 2\epsilon^2 \label{eggl:eq:emaxeav}
\end{equation}
where
\begin{equation}
\epsilon=\frac{5}{4}\frac{a_p}{a_b}\frac{e_b}{1-e_b^2}.
\end{equation}
For dynamically less excited states, i.e., $e_p(0)=\epsilon$, we have
\begin{equation}
e_p^{max}=\epsilon, \qquad \langle e^2_p\rangle=\epsilon^2 \label{eggl:eq:emaxeavef}.
\end{equation}
A word of caution: Eccentricity estimates such as the one above are based on secular orbit evolution theory. As they lack short period and resonant terms, they may not always provide accurate estimates.
More elaborate estimates can be found in \citet{eggl-et-al-2012,georgakarakos-2015,andrade-2016,georgakarakos-2016}.
Dynamically informed habitable zones for the $\alpha$ Centauri system are presented in Figure \ref{eggl:fig:dihz-s}. The difference in the permanent and averaged habitable zones suggests a strong dependence of the extent of the habitable region on the climate inertia of a potentially habitable planet. Worlds that cannot effectively buffer variations in incoming radiation would not be able to retain liquid water near their surface over roughly 50\% of the classical habitable zone around $\alpha$ Cen A. The situation is similar for potentially habitable planets orbiting $\alpha$ Cen B. If $\alpha$ Centauri had a higher orbital eccentricity, its circumstellar habitable zones would become dynamically unstable.

\subsection{Circumbinary Habitable Zones}
The maximum insolation configuration in P-type systems occurs when the planet comes closest to the brightest star.
The minimum insolation configuration is reached when the brightest star is farthest from the planet.
Hence, we can derive permanent habitable zone limits via
\begin{eqnarray}
PHZ_I&:\quad \frac{\mathbb A_I} {(q_p - \mu Q_b)^2}+\frac{\mathbb B_I}{(q_p+(1-\mu)Q_b)^2} \le 1, \label{eggl:eq:phzp2}\\
PHZ_O&:\quad \frac{\mathbb A_O} {(Q_p + \mu Q_b)^2}+\frac{\mathbb B_O}{(Q_p-(1-\mu)Q_b)^2} \ge 1, \nonumber
\end{eqnarray}
where $\mu=m_B/(m_A+m_B)$ and $\mathbb A_I>\mathbb B_I$. Here, as in S-type systems, the pericenter ($q_p$) and apocenter ($Q_p$)
distances of the planet evolve with time. The maximum insolation is, thus, always related to the maximum in the orbital eccentricity attained by the planet with respect to time.
To explicitly calculate the borders, Equation (\ref{eggl:eq:phzp2}) can be solved numerically for $a_p$.
The corresponding analytic estimates for small planetary orbital eccentricities read
\begin{eqnarray}
PHZ_I&\approx \frac{\mathbb A_I a_p} {(q_p - \mu Q_b)^2}+\frac{\mathbb B_I a_p}{(q_p+(1-\mu)Q_b)^2}, \label{eggl:eq:phz3}\\
PHZ_O&\approx \frac{\mathbb A_O a_p} {(Q_p + \mu Q_b)^2}+\frac{\mathbb B_O a_p}{(Q_p-(1-\mu)Q_b)^2}, \nonumber
\end{eqnarray}
with $q_p=c_{I}=\sqrt{\mathbb A_{I}+\mathbb B_{I}}$ and $a_p=q_p/(1-e_p^{max})$ for the inner border of the circumbinary habitable zone, and $Q_p=c_{O}=\sqrt{\mathbb A_{O}+\mathbb B_{O}}$ where $a_p=Q_p/(1+e_p^{max})$ for the outer border. Maximum eccentricities are evaluated at $q_p$ and $Q_p$ for the inner and outer border, respectively.
While Equations~(\ref{eggl:eq:emaxeav}) and (\ref{eggl:eq:emaxeavef})
still hold for $e_p^{max}$ and $\langle e_p^2 \rangle$, the expression for the forced eccentricity $\epsilon$ in circumbinary systems is different from that in circumstellar systems. Following \citet{morikawi-2004} we find that
\begin{equation}
\epsilon= \frac{5}{4}\frac{a_b}{a_p} (1-2\mu) \frac{4 e_b+3 e_b^3}{4+ 6 e_b^2},
\end{equation}
for circumbinary planets orbiting in the same plane as the stars.
To calculate circumbinary averaged habitable zone borders, we make use of equivalent radii.
Defining
\begin{eqnarray}
\bar r_p:=a_p (1-\langle e_p^2 \rangle)^{1/4},
&\bar r_{bA}:=\mu a_b(1-e_b^2)^{1/4},
&\bar r_{bB}:=(1-\mu) a_b(1-e_b^2)^{1/4},
\end{eqnarray}
the insolation averaged over a planetary orbit becomes
\begin{eqnarray}
\left \langle \mathbb S \right \rangle_t &\approx& \frac{1}{2\pi}\int_0^{2\pi}\left( \frac{\mathbb A}{\bar r_{bA}^2 + \bar r_p^2+2 \bar r_{bA} \bar r_p \cos \Phi}+\frac{\mathbb B}{\bar r_{bB}^2+\bar r_p^2-2  \bar r_{bB} \bar r_p \cos \Phi}\right)\; d\Phi,\\
 &=&\frac{\mathbb A}{\bar r_p^2 - \bar r_{bA}^2}+\frac{\mathbb B}{\bar r_p^2 - \bar r_{bB}^2}.
\end{eqnarray}
The circumbinary averaged habitable zone then reads
\begin{eqnarray}
AHZ_{I,O}&\approx \frac{\mathbb A_{I,O}}{\bar r_p^2 - \bar r_{bA}^2}+\frac{\mathbb B_{I,O}}{\bar r_p^2 - \bar r_{bB}^2}.
\end{eqnarray}

\begin{figure}[H]
\centering
%\begin{tabular}{ll}
\includegraphics[scale=.73]{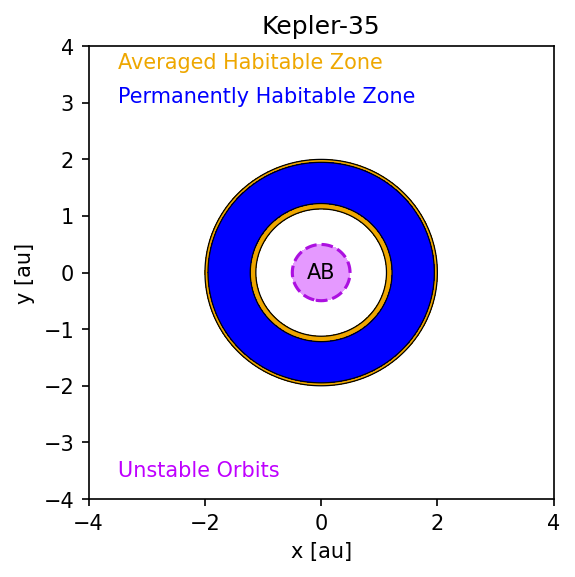}
\includegraphics[scale=0.9]{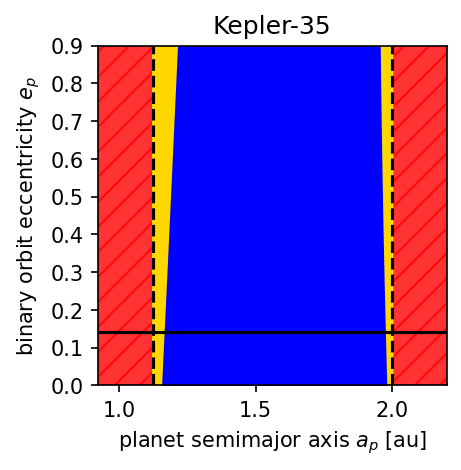}
%\end{tabular}
\caption{Same as Figure \ref{eggl:fig:dihz-s}, only for a system similar to Kepler-35. This system contains only one Earth-like circumbinary planet.
\label{eggl:fig:dihz-p}}       % Give a unique label
\end{figure}

Figure \ref{eggl:fig:dihz-p} shows dynamically informed habitable zones for a Kepler-35-like system. We assume that no other planets except a potentially habitable world are present. Circumbinary habitable zones in systems with similar stars do not depend as strongly on planetary climate inertia as S-type systems. This is evident form the small difference in the extent of permanently and averaged habitable zones for a broad range of stellar orbital eccentricities and a consequence of the fact that the forced eccentricity vanishes for mass ratios $\mu \approx0.5$. Different stellar types in a binary, however, will cause the habitability of such systems to become more dependent to the climate inertia of the planet.

%%%%%%%%%%%%%%%%%%%%%%%%%%%%%%%%%%%%%%%%%%
\section{Self-Consistent Models}
\label{eggl:sec:selfc}
Combining analytic insolation estimates with precomputed
spectral weights allows for a quick assessment of where habitable worlds can be expected in binary star systems.
This approach does have its limits, however.
Orbit evolution models based on the three body problem, for instance, do~not account for additional perturbers. Other planets in the system can alter the orbit of a potentially habitable planet. More complete dynamical models are required to account for such effects
\citep{bazso-2016,forgan-2016}.
Precomputed climate collapse criteria, such as runaway greenhouse or freeze-out limits of atmospheric greenhouse gases, may not always accurately reflect the response of a planetary climate to insolation forcing, either.
Self-consistent simulations of climate and orbital evolution of a planet are more suitable to study such phenomena in detail including resonant responses.
By coupling orbit propagators to climate models fully self consistent approaches can provide a more detailed picture
of the climate evolution of a planet in binary star systems. Such examples are the works of \citet{moorman-2019,forgan-2016,popp-eggl-2017}, which used 1D climate models, longitudinally averaged energy balance models (LEBMs)
and general circulation models (GCMs), respectively.
Self-consistent habitable zone calculations are time-consuming, however, and tuned to a specific climate model. Hence, results have to be interpreted with care \citep{forgan-2013,popp-eggl-2017}.

\section{Comparing Habitable Zones}

Comparing single star habitable zones, radiative habitable zones and dynamically informed habitable zones for the actual $\alpha$ Centauri
system, as~well as several circumbinary systems we find that analytically derived habitable zone estimates tend to coincide for systems with planets that can buffer insolation variations to a high degree.
This can be seen in Table \ref{eggl:tab:hzs} when comparing radiative habitable zone to averaged habitable zone limits.
Since perturbations on the orbit of the planet do not impact its habitability significantly in such cases,
even results obtained from single star approximations differ very little from radiative habitable zone and averaged habitable zone values.
On the other hand, if potentially habitable worlds have a lower climate inertia,
which~is likely the case near the outer rim of the single star habitable zone
\citep{popp-eggl-2017}, habitable zones in binary star systems could be substantially smaller.
This can be seen in Figures \ref{eggl:fig:dihz-s} and \ref{eggl:fig:dihz-p} when the permanent habitable zone is compared to the averaged habitable zone.
More details on this subject can be found in \citet{eggl-2018}.

\begin{table}[H]
\caption{Habitable zone borders for the $\alpha$ Centauri and Kepler-35 system.
All habitable zone values are given in [au].
The habitable zone limits for S-type A systems are given with respect to star A,
and for S-type B systems with respect to star B, respectively. Circumbinary habitable zone borders are given with respect to the barycenter of the binary.
Permanently habitable zones (PHZ) are derived assuming that the planet started on an initially circular orbit, whereas $PHZ^*$ values are derived for planets on orbits with forced eccentricity ($e_p=\epsilon$). $^{(i)}$ the corresponding habitable zone borders may be affected by orbital instability. $^{(*)}$
In the case of Kepler-35 the single star habitable zone is derived using the combined flux of Kepler-35 A and B originating
from the barycenter of the system. Stellar parameters for Kepler-35 A and B are given in Table \ref{eggl:tab:stars}.
The gravitational effect of the exoplanet Kepler-35ABb has been neglected.
$^{(**)}$ same configuration as in Figure \ref{eggl:fig:ptype-iso} right panel $a_b=0.5$ au, $e_b=0$, $^{(***)}$ same configuration as above only with $e_b=0.5$.}
\label{eggl:tab:hzs}       % Give a unique label
\begin{tabular}{p{1.9 cm}p{1.2 cm}p{0.9cm}p{0.9cm}p{0.8cm}p{0.8cm}p{0.8cm}p{0.8cm}p{0.8cm}p{0.8cm}p{0.8cm}p{0.8cm}}
\hline\noalign{\smallskip}
System & Type & SSHZ$_I$ & SSHZ$_O$ & RHZ$_I$ & RHZ$_O$ & PHZ$_I$ & PHZ$_O$ & PHZ$_I^*$ & PHZ$_O^*$ & AHZ$_I$ & AHZ$_O$ \\
%\noalign{\smallskip}\svhline\noalign{\smallskip}
\hline
$\alpha$ Centauri & S-type A & 1.17   & 2.06 & 1.18 & 2.09 &  1.29  & 1.79 & 1.23    &  1.92$^i)$ & 1.18 & 2.13$^i)$   \\
$\alpha$ Centauri & S-type B & 0.69   & 1.24 & 0.72 &1.32  & 0.74   & 1.14 &  0.72   &1.19   & 0.71 & 1.29\\
Kepler-35$^{*)}$ 	& P-type 	& 	1.12  & 1.99	& 1.16 &  1.96  &	1.23 &   1.95 & 1.23 &  1.95&  1.15 & 2.01 \\
Figure 2$^{**)}$ & P-type & 1.37 & 2.42 &    1.50 &        2.54&   1.69      & 2.31  &      1.69  &      2.31  &      1.58 &     2.53 \\
Figure 2$^{***)}$ & P-type & 1.37 & 2.42  &  1.43 &   2.62&   2.07  &  2.26 &  1.97  &  2.31 &   1.55  &  2.51 \\
\hline
\end{tabular}
%Give details in a table foot note. $^a$ This is a comment to an entry in the table
\end{table}

\section{Summary and Conclusions}
The aim of this article was to grant the inclined reader some insight into the various species of habitable zones in binary star systems.
Understanding the rationale behind different concepts is crucial in choosing which concept is best suited for predictions of where to look for
habitable worlds in a multi star environment.
Isophote-based habitable zones may be one of the most readily accessible concepts, but care has to be taken in interpreting the results. A large isophote-based habitable zone, for instance,
does not necessarily imply that a system has a greater chance of hosting habitable worlds. An increase in isophote-based habitable zones compared to classical circumstellar habitable zones can be counteracted by dynamical instability and gravitational perturbations distorting the orbit of the planet.
The combination of radiative habitable zone and orbital stability provides a better framework that only starts to break down when gravitational perturbations on planets with a low climate inertia become non-negligible.
Dynamically informed habitable zones can deal with the later cases, but they require more knowledge of the properties of the system. Dynamically informed habitable zones based on analytic estimates presented in this work start to become inaccurate when resonances come into play. Full scale simulations can be used to study resonant phenomena in climate and orbital dynamics of exoplanetary systems. However, detailed simulations with climate models are computationally expensive and it can be difficult to generalize results.
We, thus recommend that the model complexity is adapted to the specific~use case.

%%%%%%%%%%%%%%%%%%%%%%%%%%%%%%%%%%%%%%%%%%
%%%%%%%%%%%%%%%%%%%%%%%%%%%%%%%%%%%%%%%%%%
\section*{Open Source Software}
All materials, data, and computer code associated with this article are publicly accessible upon request to the corresponding author. Python codes for calculating and visualizing dynamically informed habitable zones are publicly available at \url{https://github.com/eggls6/dihz}.

%%%%%%%%%%%%%%%%%%%%%%%%%%%%%%%%%%%%%%%%%%
% \authorcontributions{Conceptualization, S.E. and E.P-L.; methodology, S.E.; software, S.E.; validation, S.E. N.G. and E.P-L.;  investigation, S.E.; resources, S.E. N.G. and E.P-L.; data curation, S.E.; writing--original draft preparation, S.E.; writing--review and editing, S.E. N.G. and E.P-L.; visualization, S.E.; supervision, S.E.; project administration, E.P-L.; funding acquisition, S.E. and E.P-L. }

%%%%%%%%%%%%%%%%%%%%%%%%%%%%%%%%%%%%%%%%%%
% \funding{``This research was in part funded by the FWF Austrian Science Fund project S11603-N16, EPL from project S11608-N16''. ``The APC was funded by E.P-L.''}

%%%%%%%%%%%%%%%%%%%%%%%%%%%%%%%%%%%%%%%%%%
\section*{Acknowledgments}S.E. acknowledges support from the DiRAC Institute in the Department of Astronomy at the University of Washington. The DIRAC Institute is supported through generous gifts from the Charles and Lisa Simonyi Fund for Arts and Sciences, and the Washington Research Foundation. This research was in part funded by the FWF Austrian Science Fund project S11603-N16, EPL from project S11608-N16

%%%%%%%%%%%%%%%%%%%%%%%%%%%%%%%%%%%%%%%%%%
% \conflictsofinterest{``The authors declare no conflict of interest.''}

%%%%%%%%%%%%%%%%%%%%%%%%%%%%%%%%%%%%%%%%%%
%% optional
\clearpage
\section*{Abbreviations and Symbols}
The following abbreviations are used in this manuscript:

\noindent
\begin{tabular}{@{}ll}
AHZ & Averaged Habitable Zone\\
HZ & Habitable Zone\\
IHZ & Isophote-based Habitable Zone\\
PHZ & Permanently Habitable Zone\\
RHZ & Radiative Habitable Zone \\
SSHZ & Single Star Habitable Zone\\
$a$ & distance between the planet and star A\\
$\mathbb A$ & spectrally weighted insolation of star A \\
$a_b$ & orbital semi-major axis of binary star\\
$a_p$ & orbital semi-major axis of the planet\\
$b$ & distance between the planet and star B\\
$\mathbb B$ & spectrally weighted insolation of star B \\
$c$ & distance of circumbinary planet to the center of reference\\
$d$ & distance between the two stars\\
$\delta$ & semi-distance between the two stars\\
$e_b$ & orbital eccentricity of the binary star\\
$e_p$ & orbital eccentricity of the planet\\
$\epsilon$ & forced orbital eccentricity of the planet\\
$f_p$ & true anomaly of the planet\\
$\phi$ & angle between the vectors connecting the two stars and the planet\\
$\mu$ & stellar mass ratio\\
$L$ & luminosity\\
$m$ & mass\\
$q_p$ & pericenter distance of the binary\\
$q_p$ & pericenter distance of the planet\\
$Q_b$ & apocenter distance of the binary star\\
$Q_p$ & apocenter distance of the planet\\
$R$ & stellar radius\\
$r$ & distance of planet to its host star \\
$r_p$ & distance of planet to the focus of the orbit\\
$\bar r_p$ & equivalent radius for the planet\\
$\bar r_b$ & equivalent radius for the binary\\
$\mathbb S$ & combined spectrally weighted insolation on the planet\\
$T$ & reduced temperature \\
$T_{eff}$ & stellar effective temperature\\
\end{tabular}

%=====================================
% References, variant B: external bibliography
%=====================================
%\externalbibliography{yes}
\bibliographystyle{abbrvnat}
\bibliography{bhz}

%%%%%%%%%%%%%%%%%%%%%%%%%%%%%%%%%%%%%%%%%%
%% optional
%\sampleavailability{Samples of the compounds ...... are available from the authors.}

%% for journal Sci
%\reviewreports{\\
%Reviewer 1 comments and authors’ response\\
%Reviewer 2 comments and authors’ response\\
%Reviewer 3 comments and authors’ response
%}

%%%%%%%%%%%%%%%%%%%%%%%%%%%%%%%%%%%%%%%%%%
\end{document}